\documentclass[pra,superscriptaddress, twocolumn, showpacs, nofootinbib]{revtex4}
\usepackage{graphicx}          % for eps figures
\usepackage{amsfonts}

\begin{document}

\title{Cloning and Joint Measurements of Incompatible Components of Spin}
\author{Thomas Brougham, Erika Andersson and Stephen M. Barnett}
\affiliation{SUPA, Department of Physics, University of Strathclyde, Glasgow G4 0NG, UK}
\date{\today}
\pacs{ 03.67.-a, 03.65.Ta}

\begin{abstract}
A joint measurement of two observables is a {\it simultaneous} measurement of both quantities upon the {\it same} quantum system.  When two quantum-mechanical observables
do not commute, then a joint measurement of these observables cannot be accomplished by projective measurements alone. In this paper we shall discuss the use of quantum
cloning to perform a joint measurement of two components of spin associated with a qubit system. 
We introduce a cloning scheme which is optimal with respect to this task.  This cloning scheme may be thought to work by cloning two components of spin onto its outputs. We
compare the proposed cloning machine to existing cloners.

\end{abstract}
\maketitle

\section{Introduction}
Quantum information has some fundamental differences from classical information.  One of the most famous of these is the inability to perfectly clone an arbitrary
quantum state.  This important observation, first enunciated by Wootters and Zurek, is referred to as the no-cloning theorem \cite{noclone}.  While quantum mechanics
precludes perfect cloning it does allow one to create approximate clones.  The first such scheme was implicit in the proof of the no-cloning theorem.  This cloning
procedure perfectly cloned states from a designated orthogonal basis, but faired less well with states that were a superposition within the prescribed basis.  

Since this work many other cloning machines have been devised.  Although the various cloners all share the same common goal of trying to copy a quantum state, the precise way that this is achieved differs due to the different specifications that they are subject to.  One example is the 
{\it universal cloner} of Buzek and Hillery \cite{universal}, which was designed to clone all qubit states equally well and   
produces two identical output cloned
states. Dropping the requirement that the clones should be identical is sometimes useful and leads to the {\it universal asymmetric cloner} \cite{cloning4, cloning5}.  Another approach 
is to try to clone a restricted set of states as well a possible. Such cloners are known as
state dependent cloners \cite{cloning1}.  The cloning machine of Wootters and Zurek is an example of a state dependent cloner, where the states of interest are orthogonal.  The examples
given thus far have applied to cloning 2-level systems (qubits); imperfect cloning of $n$-level systems is, however, also possible \cite{cloning6, cloning7}.  Yet another important development
is the experimental implementation of cloning \cite{experiment1, experiment2, experiment3}.  For a thorough review of all these topics see \cite{cloningRev}.

The cloners described thus far are sometimes referred to as {\it deterministic}, as they always return the same output for each particular input state. {\it Probabilistic
cloners} also exist \cite{cloning2, cloning3}. These allow a state, drawn from a specific linearly independent set, to be cloned exactly with a certain probability, in such a way that one knows when the  cloning has succeeded and when it has failed.
%A tentative analogue 
The relationship between deterministic and probabilistic cloners is similar to the relationship between minimum error state discrimination and unambiguous state discrimination
\cite{cloning3}.  In minimum error state discrimination, we are given a quantum state drawn from a known set of states, and we seek to determine which one.   
If, however, the states we wish to distinguish between are non-orthogonal, then we cannot distinguish them perfectly.  
Therefore we aim to distinguish between the states as well as possible, by minimising the overall probability of error.  In unambiguous state discrimination, on the other
hand, we seek to never mis-identify a state. The price we must pay to achieve this is to accept that we will sometimes not get an answer at all. To illustrate this, consider
the case of two non-orthogonal states $\{|s_1\rangle$, $|s_2\rangle\}$. For unambiguous discrimination we would require three outcomes corresponding to the state being $|s_1\rangle$, $|s_2\rangle$, and to the outcome of the experiment being undetermined.  For a review of state discrimination see \cite{chefles}.  %This analogue has been investigated in \cite{cloning3}.  

In this paper we shall explore another connection between measurement and cloning. One way of performing a {\it joint measurement} of two observables would be to first clone the
quantum system to produce two copies, and then to measure one quantum observable on each clone.  While this approach can be used to jointly measure any pair of observables, it is
important to note that current cloning procedures have not been devised with this in mind. We will derive the optimal cloning procedure with respect to measuring two incompatible
spin components of a spin-1/2 particle. 
The paper will be organised in the following way: In section \ref{sharpness}, we introduce joint quantum measurements. 
We evaluate how effective current cloners are for the task of performing a joint measurement of two spin components of a spin-1/2 particle.  After this we shall describe an existing scheme
for performing a joint measurement of two components of spin.  This will lead us, in section \ref{construct}, to a cloning scheme which is optimal with respect to the task
of performing a joint measurement of spin.  In section \ref{fid} we shall calculate the global fidelity for this cloning procedure.  We conclude with a discussion in Section
\ref{discussion}.

\section{%Sharpness of a 
Joint quantum measurements}
\label{sharpness}
A joint quantum measurement of two observables is a {\it single} measurement of a quantum system  that allows us to simultaneously give values to {\it both} observables. 
When the two observables of interest commute, then the  
joint measurement can be accomplished with standard von Neumann or projective quantum measurements.   
When the observables do not commute then we must adopt a more generalized view of quantum measurements.  
This is provided by the probability operator measures (POMs), also called positive operator valued measures (POVMs).  In the POM formalism each
measurement outcome has assigned to it a measurement operator $\hat\Pi_i$, with the probability for outcome $i$ given
by $\text{Tr}\{\hat\rho\hat\Pi_i\}$ for any measured state $\hat\rho$. This requirement leads to the fact that the measurement operators $\Pi_i$ must have eigenvalues which are
either positive or zero.  There is no requirement, however, that the measurement operators are projectors onto eigenstates. The measurement operators must also sum to the
identity operator, as the sum of all the probabilities for different outcomes is one. 
We should point out that one feature of generalized measurments is that the number of measurement operators and outcomes is not restricted to be less than or equal to the number of dimensions of the measured quantum system.
A detailed discussion of this elegant approach can be found in references \cite{nch, preskill, peres}.  

One of the earliest investigations of joint measurements was that of Arthurs and Kelly \cite{AK}, who discussed simultaneous measurements of position and momentum.  Their method
was to introduce two ancillary systems that would track both the position and momentum of the particle.  This method was extended by Arthurs and Goodman to yield an uncertainty principle for any two jointly measured observables \cite{AG}.

It is frequently assumed that joint measurements satisfy the {\it joint unbiasedness condition} \cite{AG, erikas, hall}\footnote{It is not necessary to assume the joint unbiasedness
condition.  Relaxing this condition would lead to a more general description of joint measurements \cite{hall}.}. 
This condition states that the expectation values for the jointly measured observables should be proportional to the expectation values for the observables measured by
themselves.  For the case of jointly measuring two components of a spin-$1/2$ system, we shall denote the directions of the two components by the unit vectors ${\bf a}$ and
${\bf b}$.  The observables that we seek to measure are $\hat A={\bf a\cdot\hat\sigma}$ and $\hat B={\bf b\cdot\hat\sigma}$, where ${\bf\hat\sigma}$ is a vector, the
cartesian components of which are the familiar Pauli matrices.  The  
joint unbiasedness condition will now take the form $\langle\hat A_J\rangle=\alpha\langle \hat A\rangle$ and $\langle\hat B_J\rangle=\beta\langle \hat B\rangle$, where
$\langle\hat A_J\rangle$ and $\langle\hat B_J\rangle$ are the expectation values for the jointly measured observables.  We will assume that the measurement outcomes are $\pm 1$,
so that the magnitude
of the real constants $|\alpha|$ and $|\beta|$ will vary between one and zero.  If they assume the value one, then we say that the measurement of the associated
component is completely sharp, and it will correspond to a projective or von Neumann measurement of the component.  Alternatively if one of the constants is zero then the measurement
of the associated component is said to be completely unsharp and we would have done no worse by guessing the outcome.  

The price for performing a joint measurement of incompatible observables is that the uncertainty in the %outcomes of the measurement will increase.  
estimates of the observables will increase. In other words, the variances of jointly measured spin components will be larger than the variances of the spin components measured by themselves.  This leads us to say %to the terminology 
that joint measurements are unsharp \cite{opQM}.  For the case of a spin-1/2 particle,
the values of $\alpha$ and $\beta$ will be restricted by the inequality \cite{busch}
\begin{equation}
\label{ineq}
 |\alpha{\bf a}+\beta{\bf b}|+|\alpha{\bf a}-\beta{\bf b}|\le 2.
\end{equation}
If a joint measurement scheme allows us to saturate inequality (\ref{ineq}), then, for given directions {\bf a} and {\bf b}, this measurement gives 
the largest possible values of  $\alpha$ for a given  $\beta$.  Any joint measurement of two components of spin  for which inequality  
(\ref{ineq}) is saturated shall be called an {\it optimal} joint measurement.

We can now evaluate how effective the various existing cloners are for performing joint measurements. It turns out that none of these  
allow us to saturate the inequality (\ref{ineq}) for all states, as they have not been devised  with joint measurements in mind. As an example we will consider the universal
cloner \cite{universal}. This cloner produces two identical cloned states, each having a Bloch vector pointing in the same direction as the original Bloch vector of the state. The magnitude of
the Bloch vectors is, however, reduced by the factor ${2\over 3}$.  Hence if we initially had the state $\hat \rho={1\over 2}(\hat 1+{\bf c}\cdot\hat\sigma)$, then the universal
cloner would return two identical states
$\hat \rho_c={1\over 2}(\hat 1+{2\over 3}{\bf c}\cdot\hat\sigma)$.  If we measure the ${\bf a}$ component of spin on one clone, and the ${\bf b}$ component of spin on the  other
clone, then we can realise a joint measurement of the two components with $\alpha$ and $\beta$ both equal to ${2\over 3}$.  It can be seen that the inequality (\ref{ineq}) can
never be saturated for these values of the $\alpha$ and $\beta$.  Even if we use a cloner that operates under less restrictive conditions, such as the universal asymmetrical cloner
\cite{cloning4, cloning5}, (\ref{ineq}) cannot be satisfied except in the trivial cases of $\alpha=0$, $\beta=1$ or $\alpha=1$, $\beta=0$.  This begs the question of whether it
is possible to construct a cloning machine that can be used to perform an optimal joint measurement.  In section \ref{construct}, we shall show this can be done.

\begin{figure}
\center{\includegraphics[width=9cm,height=!]
{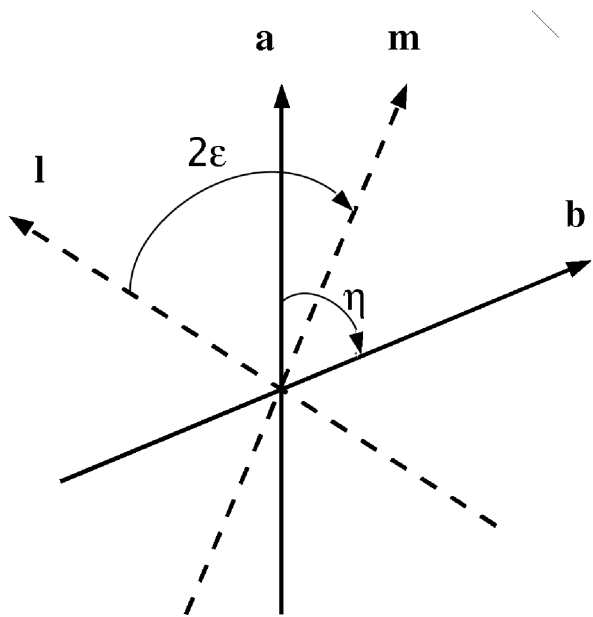}}
\caption{A diagram showing the ${\bf m}$ and ${\bf l}$ axes in relation to the ${\bf a}$ and ${\bf b}$ axes.}
\label{fig1}
\end{figure}

\section{an optimal joint measurement}
\label{realisation}
As has just been stated, none of the existing cloning machines can be used to perform an optimal joint measurement.  However such measurement schemes do exist, and a particularly
simple one is outlined in \cite{busch2, demuynck, steves, erikas}. The essence of this procedure is to introduce two new axes to measure along, which are shown in Fig. \ref{fig1} as ${\bf m}$ and ${\bf l}$. 
Each time a measurement is performed, a choice of which axis to measure along is made. The probability of measuring along ${\bf m}$ is given by $p$ and the probability of
measuring along ${\bf l}$ by $1-p$.  If we measure along ${\bf m}$ and obtain the outcome spin up along ${\bf m}$, then this is interpreted as spin along both ${\bf a}$ and ${\bf
b}$ being spin up. Conversely if we obtain spin down along ${\bf m}$, then this is interpreted as spin down along both ${\bf a}$ and ${\bf b}$.  For the case when we measure
along ${\bf l}$, the result spin up along ${\bf l}$ is taken as the component of spin along ${\bf a}$ being spin up and the component of spin along ${\bf b}$ being spin down, and
the result spin down along ${\bf l}$ is interpreted as ${\bf a}$ being spin down and the ${\bf b}$ being spin up.  

It can be seen that the four measurement operators that describe the measurement are 
\begin{eqnarray}
\label{mo}
\hat\Pi^{ab}_{\pm\pm}&=&\frac{p}{2}(\hat 1\pm{\bf m}\cdot\hat\sigma), \nonumber \\
\hat\Pi^{ab}_{\pm\mp}&=&\frac{1-p}{2}(\hat 1\pm{\bf l}\cdot\hat\sigma).  
\end{eqnarray}
The joint probability distribution $P^{ab}_{ij}$ can be calculated by taking the expectation values of the measurement operators (\ref{mo}).  The marginal probability
distributions, $P^{\alpha a}_i$ and $P^{\beta b}_j$, can then be obtained from the joint probability distribution.  Equivalently, we may obtain the marginal probability distributions directly from the marginal measurement operators $\hat\Pi^{\alpha{\bf a}}_\pm=\hat\Pi^{ab}_{\pm\pm}+\hat\Pi^{ab}_{\pm\mp}$ and $\hat\Pi^{\beta{\bf
b}}_\pm=\hat\Pi^{ab}_{\pm\pm}+\hat\Pi^{ab}_{\mp\pm}$.  

The directions of ${\bf m}$ and ${\bf l}$ can be deduced from the condition of joint unbiasedness.  This criterion imposes the constraint $\langle\hat A_J\rangle=\alpha\langle\hat A\rangle$, which implies that 
\begin{eqnarray}
\label{ju1}
\langle\hat A_J\rangle= \langle\hat\Pi^{\alpha{\bf a}}_+\rangle-\langle\hat\Pi^{\alpha{\bf a}}_-\rangle&=&\alpha\langle{\bf a}\cdot\hat\sigma\rangle, \nonumber\\
\iff \langle p{\bf m}\cdot\hat\sigma\rangle +\langle(1-p){\bf l}\cdot\hat\sigma\rangle&=&\alpha\langle{\bf a}\cdot\hat\sigma\rangle.
\end{eqnarray}
Similarly, the constraint %on $\hat B$, 
$\langle\hat B_J\rangle=\beta\langle\hat B\rangle$ implies that
\begin{equation}
\label{ju2}
\langle p{\bf m}\cdot\hat\sigma\rangle-\langle(1-p){\bf l}\cdot\hat\sigma\rangle=\beta\langle{\bf b}\cdot\hat\sigma\rangle.
\end{equation}
The conditions (\ref{ju1}) and (\ref{ju2}) suggest that a suitable choice for ${\bf m}$ and ${\bf l}$ would be
\begin{eqnarray}
\label{m&l}
{\bf m}=\frac{1}{2p}(\alpha{\bf a}+\beta{\bf b}), \nonumber \\
{\bf l}=\frac{1}{2(1-p)}(\alpha{\bf a}-\beta{\bf b}). 
\end{eqnarray}
As these are unit vectors, it can be seen that
\begin{eqnarray}
\label{prob}
p={1\over 2}|\alpha{\bf a}+\beta{\bf b}|, \\
1-p={1\over 2}|\alpha{\bf a}-\beta{\bf b}|,
\end{eqnarray}
and thus the condition that $p+(1-p)=1$ ensures that this measurement scheme is optimal.  A more detailed discussion can be found in \cite{erikas}.  This measurement
procedure can be used
to construct a cloner that enables us to perform an optimal joint measurement, which will be considered next.

\section{Construction of Cloner}
\label{construct}
It should not be surprising that the current cloning procedures do not represent optimal means of realising joint measurements.  This is as they have been conceived with the
notion of maximising the fidelity between the input state and the output cloned states.  The cloning device that we shall outline is designed to saturate inequality (\ref{ineq}), and as
such it should be expected that its fidelity will be lower than that of some of the existing cloners.  Heuristically we can imagine that this cloner operates by cloning the ${\bf
a}$ and ${\bf b}$ components of spin of the input state onto its outputs.  Now let $|a\pm\rangle$ and $|b\pm\rangle$ represent the eigenstates of spin along ${\bf a}$
and ${\bf b}$.  In addition to this let $|\psi\rangle$ represent the state that is to be cloned and let $|0\rangle$ represent the second `blank'
qubit to which information will be transferred .  Then we should expect that the action of our cloner is
\begin{eqnarray}
\label{action}
\hat U|\psi\rangle_1|0\rangle_2 =|c\rangle_{12}&=&\lambda_1|a+\rangle_1 |b+\rangle_2 + \lambda_2|a+\rangle_1 |b-\rangle_2  \\
&+&\lambda_3|a-\rangle_1 |b+\rangle_2 +\lambda_4|a-\rangle_1|b-\rangle_2,\nonumber
\end{eqnarray}
where $|\lambda_1|^2=P^{ab}_{++}$, $|\lambda_2|^2=P^{ab}_{+-}$, $|\lambda_3|^2=P^{ab}_{-+}$ and $|\lambda_4|^2=P^{ab}_{--}$, where these joint probabilities are those of an optimal joint measurement.
%the same ones which are calculated from (\ref{mo}).  
Thus we find that the squares of the magnitudes
of the coefficients in (\ref{action}) yield the same joint probability distribution as we would calculate from the measurement operators (\ref{mo}).  Now a projective measurement along the ${\bf
a}$ direction on the first qubit and along ${\bf b}$ on the second qubit of $|c\rangle_{12}$ will implement a joint measurement of the ${\bf a}$ and ${\bf b}$ components of
spin of the state $|\psi\rangle$.

To construct the unitary operator $\hat U$ that implements the cloning process, we require an important result that applies to POMs.  It can be shown that any POM can be
realised as a projective measurement by extending the dimensions of the systems Hilbert space.  This result is known as {\it Naimark's theorem}, a proof of which can be
found in \cite{preskill, peres}.  To illustrate this theorem consider the pertinent situation of performing a joint measurement of two components of spin of a spin-1/2
particle.  The state $|\psi\rangle$ of the particle is described in a two-dimensional Hilbert space.  We can add another spin-1/2 particle which is prepared in the
state $|b+\rangle$.  To describe this new combined system, with state vector $|\psi\rangle_1|b+\rangle_2$, we require a four-dimensional Hilbert space.  Naimark's theorem
assures us that there exists an orthonormal basis $\{|\phi_{ij}\rangle_{12}\}$, where $i,j=\pm$, which has the property
\begin{equation}
\label{basiscondition}
|_{12}\langle\phi_{ij}|\psi\rangle_1|b+\rangle_2|^2=\langle \psi|\hat\Pi^{ab}_{ij}|\psi\rangle=P^{ab}_{ij}.
\end{equation}
Hence we may realise the joint measurement by performing a projective measurement upon the state $|\psi\rangle_1|b+\rangle_2$.  If we find such a measurement basis
$\{|\phi_{ij}\rangle_{12}\}$, then it is clear that 
\begin{eqnarray}
\label{lambdas}
|\lambda_1|=|\langle\phi_{++}|\psi\rangle|b+\rangle|,\text{   }|\lambda_2|=|\langle\phi_{+-}|\psi\rangle|b+\rangle|, \nonumber \\
|\lambda_3|=|\langle\phi_{-+}|\psi\rangle|b+\rangle|,\text{   }|\lambda_4|=|\langle\phi_{--}|\psi\rangle|b+\rangle|.
\end{eqnarray}
This observation suggests that a suitable form for $\hat U$ is\footnote{From (\ref{lambdas}) it is clear that we are free to choose the phase factors of the terms 
that appear in the $\hat U$,
without affecting the joint probability distribution for the outcomes of the joint measurement.  The choice of phases that appear in (\ref{Cloner}) was made with a view to enhancing the fidelities of the cloning process.}
\begin{eqnarray}
\label{Cloner}
\hat U=|a+\rangle_1|b+\rangle_2\langle\phi_{++}|+|a+\rangle_1|b-\rangle_2\langle\phi_{+-}|\\
-|a-\rangle_1|b+\rangle_2\langle\phi_{-+}|-|a-\rangle_1|b-\rangle_2\langle\phi_{--}|.\nonumber
\end{eqnarray}
For this unitary operator to effect the cloning procedure we would require that the state $|0\rangle_2$ in (\ref{action}) is prepared as $|b+\rangle$.  

The task of constructing the cloner has been reduced to finding a suitable orthonormal basis $\{|\phi_{ij}\rangle_{12}\}$, which is accomplished by performing a Naimark extension.  Examples of performing
such extensions may be found in \cite{preskill, atomic}.  To aid us in performing the Naimark extension we shall introduce the states $|m\pm\rangle$ and $|l\pm\rangle$ as the
eigenstates of ${\bf m}\cdot\hat\sigma$ and ${\bf l}\cdot\hat\sigma$ respectively.  Thus we may express the POM operators as $\hat\Pi^{ab}_{\pm\pm}=p|m\pm\rangle\langle m\pm|$
and $\hat\Pi^{ab}_{\pm\mp}=(1-p)|l\pm\rangle\langle l\pm|$.  It can now be shown that one choice of basis is 
\begin{eqnarray}
\label{basis}
|\phi_{++}\rangle_{12}&=&\sqrt{p}|m+\rangle_1|b+\rangle_2+\sqrt{1-p}|a+\rangle_1|b-\rangle_2, \nonumber \\
|\phi_{--}\rangle_{12}&=&\sqrt{p}|m-\rangle_1|b+\rangle_2+\sqrt{1-p}|a-\rangle_1|b-\rangle_2, \nonumber \\
|\phi_{+-}\rangle_{12}&=&\sqrt{1-p}|l+\rangle_1|b+\rangle_2 -\sqrt{p}(\cos(\epsilon)|a+\rangle_1 \nonumber \\
&+&\sin(\epsilon)|a-\rangle_1)|b-\rangle_2, \\
|\phi_{-+}\rangle_{12}&=&\sqrt{1-p}|l-\rangle_1|b+\rangle_2+\sqrt{p}(\sin(\epsilon)|a+\rangle_1\nonumber \\
&-&\cos(\epsilon)|a-\rangle_1)|b-\rangle_2, \nonumber 
\end{eqnarray}
where $\epsilon$ is half the angle between the vectors ${\bf m}$ and ${\bf l}$.  It can be verified that this basis satisfies (\ref{basiscondition}).  It may also be verified
that the basis states are orthonormal as required.  The basis (\ref{basis}) is expressed in terms of $p$ and $\epsilon$ which relate to the measurement scheme of
\cite{erikas}.  However for this cloning machine it would be more natural to express $\hat U$ in terms of $\alpha$, $\beta$ and $\eta$, the angle between ${\bf a}$ and ${\bf
b}$.  In section \ref{sharpness} we explained that $p$ can be expressed as $p={1\over 2}|\alpha{\bf a}+\beta{\bf b}|$ and $1-p={1\over 2}|\alpha{\bf a}-\beta{\bf b}|$, hence all that remains is to find how $\epsilon$ can
be expressed in terms of $\alpha$, $\beta$ and $\eta$.  This may be achieved by remembering that $2\epsilon$ is the angle between ${\bf m}$ and ${\bf l}$, and thus ${\bf
m}\cdot{\bf l}=\cos(2\epsilon)$.  From (\ref{m&l}) it is clear that
\begin{equation}
\label{epsilon}
{\bf m}\cdot{\bf l}=\frac{\alpha^2-\beta^2}{4p(1-p)}=\cos(2\epsilon).
\end{equation}

We may view the action of the cloner (\ref{action}) as taking information about the measurement statistics and transferring it into our new basis $\{|a\pm\rangle_1, |b\pm\rangle_2\}$ in such a way that the information pertaining to ${\bf a}$ being $\pm$ is associated with the basis states %that contain 
$|a\pm\rangle_1$, and likewise for ${\bf b}$ and the states $|b\pm\rangle_2$.  If we use this cloner to aid us in performing a joint measurement of the ${\bf a}$ and ${\bf
b}$ components of spin, then we would find that the probabilities for each of the four outcomes occurring is the same as for the joint measurement scheme outlined in
\cite{erikas}.  As that measurement scheme saturated the inequality (\ref{ineq}), then this implies that a joint measurement implemented using this cloning machine will also saturate (\ref{ineq}) and as such will represent an optimal joint measurement. 

An important point to note is that in general the state $|c\rangle_{12}$ will be entangled, and thus if we wished to consider only one of the cloned qubits, then the reduced
states $\hat\rho_a=Tr_2(|c\rangle\langle c|)$ and $\hat\rho_b=Tr_1(|c\rangle\langle c|)$ will be mixed states.  Thus, the lengths of the Bloch vectors ${\bf c}_a$ and ${\bf
c}_b$ for $\hat\rho_a$ and $\hat\rho_b$ respectively, will both be less than one.  It is straighforward to show that 
\begin{equation}
\label{bloch1}
{\bf a}\cdot{\bf c}_a=\alpha({\bf a}\cdot{\bf c}),
\end{equation}
where ${\bf c}$ is the Bloch vector of the initial pure state $|\psi\rangle$.  This means that the components of $\hat\rho_a$'s Bloch vector along the ${\bf a}$
axis is just the component of the original Bloch vector of the state, along ${\bf a}$, shrunk by the factor $\alpha$.  A similar result can be found for $\hat\rho_b$,
\begin{equation}
\label{bloch2}
{\bf b}\cdot{\bf c}_b=\beta({\bf b}\cdot{\bf c}).
\end{equation}
This supports the view that the cloning process imprints the individual components of spin onto its outputs.  It also shows that the expection values for $\hat A$ and $\hat B$
satisfy the joint unbiasedness condition introduced in section \ref{sharpness}.  It is interesting to examine other components of ${\bf c}_a$ and ${\bf c}_b$.  Thus let
${\bf n}$ be the unit normal vector of the plane spanned by ${\bf a}$ and ${\bf b}$, i.e. ${\bf n}\cdot{\bf a}={\bf n}\cdot{\bf b}=0$.  It can be shown that 
\begin{equation}
\label{orthocomp}
{\bf n}\cdot{\bf c}_a=\sqrt{1-\beta^2}{\bf n}\cdot{\bf c} \quad\text{  and  }\quad{\bf n}\cdot{\bf c}_b=0,
\end{equation}
and thus ${\bf c}_b$ is confined to the ${\bf a}{\bf b}$ plane. This is not entirely unexpected, as our cloning procedure is asymmetric. It ideally leaves as much information as
possible about the {\bf a} component of spin in the original qubit state, and imprints as much information as possible about the {\bf b} component of spin on the blank auxiliary
qubit. Relation (\ref{orthocomp}) shows that some of the information related to a spin component orthogonal to both {\bf a} and {\bf b} is left in the
original qubit, but none of it makes its way onto the blank qubit. This is not surprising, as we only ever intended to transfer information about the {\bf b} component onto the
blank qubit.

The cloning procedure outlined so far applies to pure states.  It is, however possible to clone mixed states using the same cloning procedure.  If we denote the state that is to
be cloned as $\hat\rho$, then the output cloned state $\hat\rho_c$ will simply be $\hat\rho_c=\hat U\hat\rho\otimes|b+\rangle\langle b+|\hat U^+$.  A joint measurement of ${\bf
a\cdot\sigma}$ and ${\bf b\cdot\sigma}$ can now be realised in the same manner as for the pure state cloning.   

\section{Fidelities}
\label{fid}
\begin{figure}
\center{\includegraphics[width=8cm,height=!]{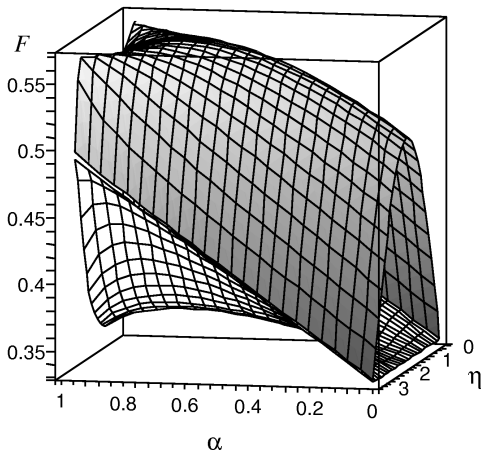}}
\caption{A plot showing $F_{av}$ (the upper surface) in relation to $F_m$ (the lower surface). The parameter $\eta$ is the angle between the $a$ and $b$ axes.}
\label{plot3}
\end{figure}

An important quantity to consider when analysing any cloning procedure is the fidelity, which quantifies how close the cloned states are to the original state.  Here we shall
take our definition of fidelity, $F$, to be
\begin{equation}
\label{fidelity}
F=|_1\langle\psi|_2\langle\psi|c\rangle_{12}|^2,
\end{equation}
which is the probability that the state produced is found, by a suitable measurement, to be a pair of perfect copies.  The fidelity (that is obtained) will depend upon the initial state and thus, as expected, it differs from the fidelity obtained using the universal cloner.  However it also
differs from the fidelities of traditional state dependent cloners, which are optimised for a restricted set of states, as this cloner is optimal with respect to (\ref{ineq})
for all pure qubit states.  It is interesting to look at how the fidelity varies with the angle between ${\bf a}$ and ${\bf b}$, and how it varies for different intended
measurement sharpness.  As the fidelity is a function of the input state, we shall average $F$ over all possible initial pure states.  This task is best accomplished in terms of
the Bloch sphere picture, where the input states are represented by unit vectors in ${\mathbb R}^3$.  Choosing ${\bf a}$ to be along the $z$-axis, we define $\theta$ to be the angle
made by the Bloch vector of the state and the $z$-axis.  We also define $\phi$ to be the angle made by the projection of the Bloch vector of the state in the $xy$ plane with the
$x$-axis.  Assuming all initial states to be equally probable, the averaged fidelity will be given by
\begin{equation}
\label{average}
F_{av}=\frac{1}{4\pi}\int^{2\pi}_0{\int^{\pi}_0{F_{(\theta, \phi)}\sin\theta d\phi d\theta}}.
\end{equation}
This averaged fidelity is equivalent to the global fidelity of \cite{cloning1}.  With some effort it can be shown that 
\begin{eqnarray}
\label{averaged}
F_{av}&=&{1\over 4}+\frac{\alpha}{12}+\frac{\beta}{12}+\frac{\alpha\beta}{12}\cos^2(\eta)+\frac{\sqrt{1-\beta^2}}{12}\nonumber \\
&+&\frac{\sqrt{1-\alpha^2}}{12}\sin(\eta)+\frac{1}{24p} [ \alpha\sqrt{1-\beta^2}+\nonumber \\
&+&\beta\sqrt{1-\beta^2}\cos(\eta)+\beta\sqrt{1-\alpha^2}\sin(\eta) ],
\end{eqnarray}
where $p$ is given by (\ref{prob}).  A plot of the averaged fidelity is shown in figure \ref{plot3}.  In this figure, %the plotting of figure \ref{fig2},
$\beta$ was chosen to have its largest possible value consistent with the
choice of $\alpha$ and $\eta$.  It was found in \cite{universal} that the universal cloner produces clones with a fidelity of ${5\over 6}$.  Thus the two particle fidelity of
the univeral cloner is $({5\over 6})^2\approx 0.6944$.  Comparing this to figure \ref{plot3} we observe that the universal cloner produces clones with a higher fidelity than
the cloner we are considering.

We can also define fidelities for the single particle reduced
states $\hat\rho_a$ and $\hat\rho_b$ to be $F_a=\langle \psi|\hat\rho_a|\psi\rangle$ and $F_b=\langle \psi|\hat\rho_b|\psi\rangle$.  These fidelities can then be averaged in the same manner as (\ref{average}).
This yields the results
\begin{eqnarray}
\label{fa}
F_a&=&{1\over 2}+{\alpha\over 6}+{1\over 6}\sqrt{1-\beta^2}+{1\over 12p}(\alpha\sqrt{1-\beta^2}+ \nonumber\\
&+&\beta\cos(\eta)\sqrt{1-\beta^2}+\beta\sin(\eta)\sqrt{1-\alpha^2}),\\
\label{fb}
F_b&=&{1\over 2}+{\beta\over 6}.
\end{eqnarray}
The fact that the averaged fidelities are not equal is to be expected, as the process that we are performing is inherently asymmetric with respect to the {\bf a} and {\bf b} directions. 
To elaborate further upon this, the reduced states $\hat\rho_a$ and $\hat\rho_b$ contain information about the measurement statistics of measurements along the ${\bf a}$ and ${\bf b}$ axes of the state $|\psi\rangle$. Provided ${\bf a}\ne{\bf b}$, we should not expect $\hat\rho_a$ and $\hat\rho_b$ to be equal, and thus we should not expect $F_a$ to equal $F_b$.  For the case when ${\bf a}={\bf b}$, i.e. $\eta=0$, it is possible to have $\alpha=\beta=1$ and thus the equations (\ref{fa}) and (\ref{fb}) will both yield the answer of $2\over 3$ for the averaged fidelity. 
The cloning process we have considered leaves information relating to spin along the {\bf a} direction in the original qubit state, and transfers information relating to spin
along {\bf b} onto a blank qubit. We could, of course, equally well consider a cloning process which would leave the information relating to {\bf b} in the original qubit, and
copy the information relating to {\bf a} onto the blank state. The fidelities in equations (\ref{fa}) and (\ref{fb}) would then be reversed with respect to $\alpha, \bf a$ and $\beta, \bf b$.
One could also consider devising a cloning procedure which would be more symmetric with respect to {\bf a} and {\bf b} as far as the fidelities for the single particle reduced states are concerned. This cloning procedure would still give the same measurement statistics. The possibility of many different cloning procedures, all yielding the same measurement statistics, is due to the fact that there are infinitely many ways of realising the joint quantum measurement, in terms of how to make the Naimark extension.

In section \ref{realisation} a way of 
realising a joint measurement of two components of spin was outlined.  As our cloner is primarily concerned with replicating measurement statistics, it may seem more natural to use the measurement scheme of section \ref{sharpness} in a more direct fashion to prepare the clones.   
We could just perform the joint measurement and then prepare one of the basis states $|a\pm\rangle_1|b\pm\rangle_2$ corresponding to the obtained outcome.  This process would, on average, yield the mixed state
\begin{equation}
\label{mixed}
\hat\rho_{12}=\sum_{i,j=+,-}{P^{ab}_{i,j}|a_i\rangle_1|b_j\rangle_{2 1}\langle a_i|_2\langle b_j|}.
\end{equation}
The state (\ref{mixed}) is simpler to create and gives the same measurement statistics as $|c\rangle$.  There is, however, a difference between the two cloning procedures.  This difference
manifests itself within the averaged fidelities of the two processes.  We can define the fidelity, $F_m$, of $\hat\rho_{12}$ with the original state
$|\psi\rangle$ as 
\begin{equation}
F_m={ }_1\langle\psi|_2\langle\psi |\hat\rho_{12}|\psi\rangle_1|\psi\rangle_2.
\end{equation}
When we average $F_m$ over all the pure qubit states, then we find that the averaged fidelity is less than (\ref{averaged}), as is shown in figure \ref{plot3}.  It is
informative to consider the two reduced states $\hat\rho^a_1=\text{Tr}(\hat\rho_{12})$ and $\hat\rho^b_2=\text{Tr}(\hat\rho_{12})$.  The averaged fidelities may be
calculated for these states and are found to be
\begin{eqnarray}
F_{ma}&=&{1\over 2}+{\alpha\over 6}, \\
F_{mb}&=&{1\over 2}+{\beta\over 6}.
\end{eqnarray}
It may be seen that $F_{mb}=F_b$ but that $F_{ma}\ne F_a$.  This result again shows that only information pertaining to the ${\bf b}$ component is transferred to the second qubit,
where as the first qubit retains additional information about the original state.

\section{Conclusion}
\label{discussion} 
We have looked at how quantum cloning can be used to perform joint quantum measurements of two components of spin.  A criterion for judging the optimality of a joint measurement
was discussed.  We then introduced a cloning machine which can be used to perform an optimal joint measurements of spin.  This cloning scheme could be though to act by cloning the ${\bf a}$ and
${\bf b}$ components of spin onto its outputs.  Fidelities for the cloner were also investigated.  We could compare these results to that of the universal cloner, which provides
clones with a fidelity of $5\over 6$ for all input states.  Thus its two particle averaged fidelity would simply be $({5\over 6})^2\approx$0.6944.  Hence the fidelity of the
universal cloner is, as expected, higher than that of the cloner presented in section \ref{construct}.  

Finally we compared the cloning procedure to a more direct approach producing the mixed state output (\ref{mixed}).  It was found that the averaged fidelity for the cloning procedure outline in section \ref{construct} was greater than or equal to the fidelity of the mixed state %other 
procedure.  Thus the added  complexity of the method outlined in section \ref{construct} may be balanced against the higher fidelity that it provides. 

\acknowledgments
We would like to thank Prof. Alain Aspect for suggesting the question of how to make a joint quantum measurement using cloning. EA acknowledges the Royal Society for financial support.


\begin{thebibliography}{00}
\bibitem{noclone} W.K. Wootters and W.H. Zurek, { Nature} (London) {\bf 299}, 802 (1982).
\bibitem{universal} V. Bu\v zek and M. Hillery, { Phys. Rev. A} {\bf 54}, 1844 (1996).
\bibitem{cloning4} V. Bu\v zek, M. Hillery and R. Bednik, { Acta Phys. Slov.} {\bf 48}, 177 (1998); e-print {quant-ph/9807086v1}, (1998).
\bibitem{cloning5} S. Ghosh, G. Kar and A. Roy, { Phys. Lett. A} {\bf 261}, 17 (1999).
\bibitem{cloning1} D. Bruss, D. P. DiVincenzo, A. Ekert, C. A. Fuchs, C. Macchiavello, J. A. Smolin, { Phys. Rev. A} {\bf 57}, 2368 (1998).
\bibitem{cloning6} R. F. Werner, { Phys. Rev. A} {\bf 58}, 1827 (1998).
\bibitem{cloning7} J. Fiur\' a\v sek, R. Filip, N. Cerf, { Quant. Inform. Comp.} {\bf 5}, 583 (2005); e-print { quant-ph/0505212v1}, (2005).
\bibitem{experiment1} C. Simon, G.Weihs and A. Zeilinger, { Phys. Rev. Lett.} {\bf 84}, 2993 (2000).
\bibitem{experiment2} F. De Martini, D. Pelliccia, F. Sciarrino, { Phys. Rev. Lett.} {\bf 92}, 067901 (2004).
\bibitem{experiment3} Z. Zhao, A. N. Zhang, X. Q. Zhou, Y. A. Chen, C. Y. Lu, A. Karlsson, J. W. Pan, { Phys. Rev. Lett.} {\bf 95}, 030502 (2005).
\bibitem{cloningRev} V. Scarani, S. Iblisdir and N. Gisin, { Rev. Mod. Phys.} {\bf 77}, 1225 (2005).
\bibitem{cloning2} L. M. Duan and G. C. Guo, { Phys. Lett. A} {\bf 243}, 261 (1998).
\bibitem{cloning3} A. Chefles, S. M. Barnett, { J. Phys. A} {\bf 31}, 10097 (1998).
\bibitem{chefles} A. Chefles, { Contemp. Phys.} {\bf 41}, 401 (2000).
\bibitem{nch} M. A. Nielsen and I. L. Chuang, Quantum Computation and Quantum Information, Cambridge University Press, Cambridge, 2000.
\bibitem{preskill} J. Preskill, Quantum Computing, Information and Complexity, lecture notes, http://www.theory.caltech.edu/people/preskill/ph229/.
\bibitem{peres} A. Peres, Quantum Theory; Concepts and Methods, Kluwer Academic Publishers, 1998.
\bibitem{AK} E. Arthurs and J. L. Kelly, { Bell Syst. Tech. J.} {\bf 44}, 725 (1965).
\bibitem{AG} E. Arthurs and M. S. Goodman, { Phys. Rev. Lett. } {\bf 60}, 2447 (1988).
\bibitem{erikas} E. Andersson, S.M.Barnett and A. Aspect, { Phys. Rev. A} {\bf 72}, 042104 (2005).
\bibitem{hall} M. J. W. Hall, { Phys. Rev. A} {\bf 69}, 052113 (2004).
\bibitem{opQM} P. Busch, M. Grabowski, P. J. Lahti, Operational quantum physics, Springer-Verlag, Berlin, 1995. 
\bibitem{busch}  P. Busch, Phys. Rev. D {\bf 33}, 2253 (1986).
\bibitem{busch2} P. Busch, { Found. Phys. } {\bf 17}, 905 (1987)
\bibitem{demuynck} W. M. de Muynck and H. Martens, Phys. Lett. A {\bf 142}, 187 (1989).
\bibitem{steves} S. M. Barnett, { Phil. Trans. Roy. Soc. Lond. A} {\bf 59}, 1844 (1996)
\bibitem{atomic} S. Franke-Arnold, E. Andersson, S. M. Barnett, and S. Stenholm, { Phys. Rev. A} {\bf 63}, 52301 (2001).

\end{thebibliography}
\end{document}